\documentclass[aps,pra,twocolumn]{revtex4}   
\newcommand {\be}{\begin{equation}}
\newcommand {\ee}{\end{equation}}
\newcommand {\bey}{\begin{eqnarray}}
\newcommand {\eey}{\end{eqnarray}}

\usepackage{epsfig}
\usepackage{amsfonts}
\usepackage{mathbbol} 

\begin{document}

\title{Compressing the hidden variable space of a qubit}

\author{Alberto Montina}
\affiliation{Perimeter Institute for Theoretical Physics, 31 Caroline Street 
North, Waterloo, Ontario N2J 2Y5, Canada}
\email{amontina@perimeterinstitute.ca}

\date{\today}

\begin{abstract}
In previously exhibited hidden variable models of quantum state
preparation and measurement, the number of continuous hidden
variables describing the actual state of single realizations is
never smaller than the quantum state manifold dimension.
We introduce a simple model for a qubit whose hidden variable
space is one-dimensional, i.e., smaller than the
two-dimensional Bloch sphere. The hidden variable probability
distributions associated with quantum states satisfy reasonable
criteria of regularity.
Possible generalizations of this shrinking to an $N$-dimensional
Hilbert space are discussed.
\end{abstract}
\maketitle

\section{Introduction}
A quantum state is not discernible by means of a single replica, but can 
be reconstructed only by performing many measurements on identically 
prepared systems~\cite{noclone}. In this sense it is analogous to a classical 
probability distribution that carries statistical information on 
the actual state of many realizations. The similarity is reinforced by the fact that both the quantum state
and the probability distribution have linear dynamical equations.
Furthermore, as for the probability distributions, the space of
quantum states of a composite system is the tensorial product of the 
subsystem Hilbert spaces. These analogies suggest the idea
that the quantum state does not represent the actual state of a
single realization, but merely contains statistical information about
an underlying hidden variable state, also named 
{\it ontic state}~\cite{spekkens}.  Indeed, a quantum state 
has the peculiar feature of containing the full statistical information 
concerning any measurement outcome performed on ensembles.
However, unlike classical mechanics,
the standard quantum framework does not provide a picture beyond this
ensemble description, which unavoidably requires an exponential growth of 
resources in the state definition. Thus, it is reasonable to wonder if the 
complexity of quantum mechanics could be mitigated by filling
the gap between classical and quantum representations. Indeed there is no 
{\it a priori} reason to believe that the ontic states must contain the full 
ensemble information contained in quantum states.

The relation between exponential complexity and ensemble description
is well-illustrated by a classical example. On the one hand, a system of $N$ 
particles is described by a set of canonical variables $\{\vec q_i,\vec p_i\}$ 
whose number scales linearly with $N$. On the other hand,
the statistical state is given by a multivariate 
probability distribution $\rho(\{\vec q_i,\vec p_i\})$, whose complete 
characterization requires generally a number of resources growing 
exponentially with $N$ for a given accuracy. This is not surprising, since
the probability distribution contains a quantity of information
that could be approached only by performing a series of measurements 
on a large number of replicas. 

Our core question is as follows. Suppose, in accordance with Einstein's 
point of view~\cite{einstein}, that there exists a more fundamental 
theory that does not use quantum states, but describes each
single system by means of well-defined (ontological) variables. 
The wave-function and the ontological variables would be analogous
to the classical multivariate probability distribution $\rho$
and the coordinates $\{\vec q_i, \vec p_i\}$, respectively.
Thus, the following question naturally arises:
can the ontological space dimension grow polynomially with
the size of the quantum system? By {\it size} here we mean
the number of elements that compose the system, for
example the number of spins. As will be shown in this
Letter, this question is anything but trivial. Indeed, we will prove with 
a practical example that the ontological space dimension can be smaller than
the quantum state manifold dimension\footnote{The projective Hilbert 
space can be considered as a manifold, i.e., it is locally homeomorphic 
to Euclidean space, its topology induced by the Fubini-Study 
metric. Its dimension is twice the Hilbert space dimension minus two.}.
We call this dimensional 
reduction {\it ontological shrinking}. The ontological shrinking is 
a necessary condition for having an exponential compression,
although not sufficient. Our result is fundamental for sweeping away 
any prejudge that this reduction is impossible.

The ontological shrinking is closely related to the concept of classical 
``weak simulation''~\cite{jozsa,denNest} in quantum information 
theory. Contrary to a ``strong simulation'' of quantum computers,
the  goal of a classical ``weak simulation'' is not
to compute the measurement probabilities with high accuracy, but 
the outcomes in accordance with the probabilities. 
There are examples of quantum circuits
that cannot be efficiently simulated in a strong way,
but whose weak simulation is nevertheless tractable~\cite{denNest}.
In a hidden variable theory with reduced sampling space,
the identification of the actual state of a single realization
would require less resources than the quantum state definition.
Thus, the ontological shrinking could offer in a natural 
way an efficient method of ``weak simulation'' of quantum computers.

The problem of the minimal quantity of ontological resources 
was recently considered  by different 
authors~\cite{hardy,montina,monti2,rudolph,brukner}. In Ref.~\cite{hardy}, 
Hardy has proved that the number of ontic states is always infinite, 
even in the case of a finite-dimensional Hilbert space. According to 
this constraint, the set of ontic states is not less than
countably infinite. That is a weak 
condition, since the Hilbert space is continuous and in general 
multi-dimensional. The problem of the smallest dimension of the 
ontological space was introduced in Ref.~\cite{montina}. It was 
subsequently proved that the ontological space dimension cannot be 
smaller than the dimension of the quantum state manifold in the case 
of a short memory hidden variable theory~\cite{monti2,monti3}, implying
an exponential growth of resources (``Short memory'' means that
the dynamics of the ontic state is Markovian). In 
Ref.~\cite{brukner} a hidden variable model for a finite number of
measurements was reported, whose number of resources saturates to 
the constraint of Ref.~\cite{monti2} in the case where the whole set of 
possible measurements is considered. 
In the conclusion of Ref.~\cite{monti2} it was noted that a possible 
way to overcome the exponential growth would be to reject one of 
the hypotheses of the theorem, e.g., the causality. Thus, the 
dynamics of the ontological state would not be described as a 
Markov succession of causes and effects. The Wheeler-Feynman absorber 
theory~\cite{wheeler} is an example of non-causal theory,
introduced in another context with the aim of
removing the self-interaction of electric charges. It is
an example of retro-causality, where future events can
have effects on the past. 
Recently, Wharton introduced a time-symmetric interpretation of
the Klein-Gordon equation by imposing boundary conditions
both in past and future~\cite{wharton}.

In this Letter, we consider the problem of describing a
process of state preparation and subsequent measurement
completely neglecting any concern about the dynamics.
We present a hidden variable model of a qubit
whose ontological manifold has one dimension, that is 
one half the dimension of the quantum state manifold
(the Bloch sphere). This is the first example of a
hidden variable model for measurements with a compressed
sampling space and raises the interesting question of whether 
an exponentially growing number of resources is in
fact required to describe the hidden variable state of
a quantum system. Furthermore, it illustrates that the 
short memory hypothesis is indeed strictly necessary
for the proof of the theorem in Ref.~\cite{monti2}.
Finally, we discuss possible generalizations to
an $N$-dimensional Hilbert space and present a model
that works for a large set of trace-one projectors.

\section{Ontological model of qubit}

In a hidden variable theory of state preparation and
measurement, the quantum state is translated into a
classical language by replacing it with a probability
distribution on a sampling space $X$ of ontic states,
$|\psi\rangle\rightarrow \rho(X|\psi)$,
i.e., it is assumed that a quantum system is
described, at a deeper and partially hidden level,
by a set of classical variables that do not necessarily
contain the whole information carried by $|\psi\rangle$.
Such information is contained in the ensemble
distribution $\rho$. 
We assume that the ontological space $X$ is a disconnected manifold
defined by a set of continuous variables $(x_1,...,x_D)\equiv\vec x$
and a discrete index $n$, labeling the components of the manifold.
Its Lebesgue covering dimension is $D$.
The distribution $\rho(\vec x,n|\psi)$ is a regular 
function, in a sense that will be specified later on.
This allows us to define the concept
of ontological space dimension, which is the
number $D$ of continuous variables.
The ontological variables are only partially hidden in 
the sense that they have a role in the generation
of an event when a measurement is performed. Let the 
trace-one projector $|\phi\rangle\langle\phi|$ be the
observable that will be measured after preparing the
state $|\psi\rangle$. At the ontological level, there is
a conditional probability $P(\phi|X)$ of obtaining 
the event $\phi$ given the ontic state $X$. The
ontological theory is equivalent to quantum mechanics if
\be\label{born} 
\sum_n\int d^D x P(\phi|\vec x,n)
\rho(\vec x,n|\psi)=|\langle\phi|\psi\rangle|^2.
\ee

It is important to note that the dimension of a manifold 
is a topological property and we have to introduce some
conditions of regularity on $\rho(X|\psi)$ in order that 
the problem of the ontological dimension is not ill-posed. 
Without these conditions the ontological shrinking would 
be a trivial task, since a multi-dimensional space has the 
same cardinality of a one-dimensional space.
Indeed, it is always possible
to construct a one-dimensional ontological model from a
multi-dimensional one in a trivial way, by means of a
bijective transformation. Let 
us consider for example the Kochen-Specker (KS) model for a 
qubit~\cite{kochen}. The corresponding ontological space is
a unit sphere and the states are unit three-dimensional
vectors $\vec s$, which can be identified by
two continuous variables $q$ and $p$, such as
the angles in the spherical coordinate system.
Each quantum state $|\psi\rangle$ is associated
with a probability distribution $\rho(q,p|\psi)$.
The mapping $|\psi\rangle\rightarrow\rho(q,p|\psi)$ in
the KS model 
is almost regular, that is, $\rho(q,p|\psi)$ is a 
differentiable function for every $|\psi\rangle$ and $(q,p)$, 
apart from a zero measure subset of ontological states.
We can rescale $q$ and $p$ in such a way that they can
be written in the decimal notation $q=0.q_1 q_2 q_3...$
and $p=0.p_1 p_2 p_3...$, $q_i$ and $p_i$ being
digits. 
The two-dimensional ontological space can be compressed
into a one-dimensional space by encoding the information 
carried by $q$ and $p$ into the single real number
$r\equiv 0.q_1 p_1 q_2 p_2 q_3 p_3...$. However, the distribution 
$\rho(r|\psi)$, with the new topological space $r$ as domain, will 
be highly irregular and physically unsuitable.

In order to rule out these trivial routes towards the
ontological shrinking, we impose the condition that the 
function $\rho(\vec x,n|\psi)$ is analytic in $\vec x$
and $|\psi\rangle$ almost everywhere, possibly
containing also a collection of Dirac functions whose 
heights and positions are differentiable functions of
$|\psi\rangle$ almost everywhere. This gives a sufficient
general condition of regularity that makes the 
ontological shrinking a non-trivial problem.

With these premises, let us consider a two-state quantum
system and introduce a one-dimensional hidden variable model 
that works only for a subset of preparation states.
Later on we will extend the model to the whole quantum
state manifold. Our ontological space is given by a
continuous variable $x$ and a discrete index $n$ that
takes the two values $0$ and $1$. It is convenient
to represent the quantum state $|\psi\rangle$ and the event
$|\phi\rangle$  by means of the Bloch vectors
$\vec v\equiv \langle\psi|\vec\sigma|\psi\rangle$ and
$\vec w\equiv \langle\phi|\vec\sigma|\phi\rangle$,
where $\vec\sigma\equiv(\hat\sigma_x,\hat\sigma_y,\hat\sigma_z)$,
the $\hat\sigma_i$ being the Pauli matrices.

The probability distribution associated with the state $\vec v$
is
\be\label{distr_qubit}
\rho(x,n|\vec v)=\sin\theta\delta_{n,0}\delta(x-\varphi)+
(1-\sin\theta)\delta_{n,1}\delta(x-\theta),
\ee
where $\varphi$ and $\theta$ are respectively the azimuth and 
zenith angles in the spherical coordinate system
\be\begin{array}{l}
v_x=\sin\theta\cos\varphi,  \\ 
v_y=\sin\theta\sin\varphi,  \\
v_z=\cos\theta. 
\end{array}
\ee
Thus, when the quantum state $\vec v$
is prepared, the index $n$ takes the value $0$ or $1$ with
probability $\sin\theta$ or $1-\sin\theta$ and the continuous 
variable is the azimuth or zenith angle according
to the value of $n$. The probability distribution $\rho(x,n|\vec v)$
is non-negative for any $\{x,n\}$ and $\vec v$.

The probability distribution~(\ref{distr_qubit}) is a collection
of two delta functions whose heights and positions are
differentiable functions of $\theta$ and $\varphi$. Thus,
they fulfil the conditions of regularity required previously.
It is important to note that a single
realization contains less information than the quantum state, 
since only one of the two angles $\theta$ and $\varphi$ is
stored in the ontic state $\{x,n\}$. 
The whole information about $\vec v$ is carried by the ensemble 
distribution $\rho(x,n|\vec v)$.

The conditional probability $P(\vec w|x,n)$ for an event $\vec w$ 
with $w_z>0$ is defined as follows:
\be\label{condprob}
\begin{array}{l}
P(\vec w|x,0)=1+\frac{w_x \cos x+ w_y \sin x-\sqrt{1-w_z^2}}{2}, \\
P(\vec w|x,1)=\frac{1+(\sqrt{1-w_z^2}-2)\sin x+
w_z\cos x}{2-2\sin x}.
\end{array}
\ee
The events $\vec w$ with $w_z<0$ correspond simply to the non-occurrence
of the events $-\vec w$ with $w_z>0$, i.e, $P(-\vec w|x,n)=1-P(\vec w|x,n)$.

It is easy to prove that these probability functions fulfil
the condition (\ref{born}), that is,
$P(\vec w|\varphi,0)\sin(\theta)+P(\vec w|\theta,1)(1-\sin\theta)=
(1+\vec w\cdot\vec v)/2$.
We have to check that the conditional probabilities satisfy
the constraints $0\le P(\vec w|x,n)\le1$. It is sufficient
to consider the case $w_z>0$. The non-negativity of
$P(\vec w|x,0)$ for any $x$ and $\vec w$ is proved by the fact 
that its minimum is zero. The minimum with a fixed $\vec w$
is equal to 
\be
1-\sqrt{1-w_z^2}\equiv m(w_z)
\ee
and taken when the vectors 
$(\cos x,\sin x)$ and $(w_x,w_y)$ are antiparallel,
that is, when $\cos x=-w_x/\sqrt{w_x^2+w_y^2}$, 
$\sin x=-w_y/\sqrt{w_x^2+w_y^2}$. The overall minimum, i.e., the
minimum of $m(w_z)$,
is taken at $w_z=0$ and is equal to $0$. Similarly, the largest value, 
with $(\cos x,\sin x)$ and $(w_x,w_y)$ parallel, is equal to $1$. The 
conditional probability $P(\vec w|x,1)$ takes its maximum when the vectors 
$(\sqrt{1-w_z^2},w_z)$ and $(\cos x,\sin x)$ are parallel and is equal 
to $1$. Its non-negativity check deserves a more detailed 
discussion. It is easy to realize that $P(\vec w|x,1)$ is negative
for some $\vec w$ if $\cos x<0$, thus this model does not work
for any state. Let us find the region with $\cos x>0$ where
the probability is non-negative for every $\vec w$. 
For a fixed $x$, the conditional 
probability is minimal at the boundary $w_z=1$ and
takes the value $\frac{1-2\sin x+\cos x}{2-2\sin x}$.
It is non-negative if
\be\label{eq_constr}
\theta=x<\theta_0\equiv\arccos(3/5)
\simeq 53.13 \text{ degrees}.
\ee

Thus, the present model works only for a set of prepared
states whose Bloch vector lies inside a cone with aperture 
$2\theta_0$, the $z$-axis being the cone symmetry axis.
It is interesting to note that there is no constraint on
$\vec w$ and the model works for any event if the prepared state
fulfils the condition~(\ref{eq_constr}).

A simple way to extend the model to any preparation state
is to increase the information contained in the ontological 
state. Let $\vec n_1,..,\vec n_{M_b}$ be $M_b$ fixed Bloch vectors.
A set of orthogonal coordinates is associated with each
$\vec n_k$. The fixed Bloch vector $\vec n_k$ is the
$z$-axis of the associated coordinate system.
When the state $\vec v$ is prepared, the information on the 
nearest $\vec n_k$ is enclosed in an additional discrete index 
$m$ that can take $M_b$ possible values. This index does not
change the ontological space dimension, which remains equal to 
one. Furthermore, 
the preparation apparatus uses the coordinate system attached 
to $\vec n_k$.
By {\it nearest vector} we mean the vector $\vec n_k$ with the
largest scalar product $\vec n_k\cdot\vec v$. 
The advantage of the added information is that the
measurement apparatus receives the information on the
closest $\vec n_k$ and can use the protocol previously 
described with the coordinate system attached to
$\vec n_k$. If the number $M_b$ of vectors $\vec n_k$ is 
sufficiently large, the angle between the closest axis and
the quantum state $\vec v$ is always smaller than
$\theta_0$. In the case of equidistributed vertices, the
smallest number $M_b$ of Bloch vectors is $12$ and they
correspond to the vertices of an icosahedron. For this
polyhedron, the length of the edges is $L=4/\sqrt{10+2\sqrt{5}}$
for a unit sphere, so the distance between the circumcenter of 
each face and its vertices is $d=L/\sqrt3$.
The angle between vectors passing through a circumcenter
and the closest vertex is $\theta_1=\arcsin d\simeq 37.37$
degrees. The inequality $\theta_1<\theta_0$ guarantees
that the angle between the $z$-axis and $\vec v$ is always
smaller than $\theta_0$. Thus, the extended ontological model 
works for any state preparation and measurement.
The patches associated with each vector are $12$ congruent spherical 
pentagons, which give a regular tessellation of the Bloch sphere. 
They are the sphere projection of dodecahedron faces. We indicate
the patch pointed to by $\vec n_k$ with $\Omega_k$.

The extended model is equivalent to the following protocol.
Suppose that Bob and Alice share a common reference frame
on the Bloch sphere and a set of Bloch vectors
$\vec n_1,..,\vec n_{M_b}$, corresponding to the
icosahedron vertices. Let the first vector $\vec n_1$
be the $z$-axis of the reference frame. Bob prepares a
quantum state $\vec v$, which for the moment is assumed to point
at the spherical pentagon $\Omega_1$ associated with $\vec n_1$.
He assigns to a discrete variable $n$ one of values $0$ and $1$, randomly 
generated with probabilities $\sin\theta$ and
$1-\sin\theta$, respectively. If $n=0(1)$, Bob sets a continuous
variable $x$ equal to the azimuth angle $\varphi$ (zenith angle $\theta$)
and sends both $n$ and $x$ to Alice. Alice generates the
event $\vec w$ with probability given by Eq.~(\ref{condprob}).
The fact that $\vec v$ points at $\Omega_1$ guarantees that
condition~(\ref{eq_constr}) is always satisfied, that is, Alice 
always receives values of $\{x,n\}$ such that $0\le P(\vec w|x,n)\le1$.
It is worth to stress that in a single
run Alice has only a partial information about the quantum
state. Indeed she knows only $\varphi$ or $\theta$, according
to the value of $n$. Nevertheless, she can generate
with this partial information events in accordance
with quantum probabilities.
The model is extended to the whole Bloch sphere as follows.
The task is generating the event $\vec w$ with probability
$(1+\vec v\cdot\vec w)/2$, given any state $\vec v$. 
Let $\vec n_k$ be the closest vector to
$\vec v$. Bob evolves the quantum state according to a unitary
evolution $\hat U_k$ that takes the spherical pentagon $\Omega_k$
to $\Omega_1$. Let $\hat O_k$ be the corresponding orthogonal
transformation on the Bloch sphere. After the transformation
$\vec v\rightarrow\hat O_k\vec v$, Bob executes the previously 
described protocol and sends to Alice the pair $\{x,n\}$ and the 
auxiliary index $k$. In order to evaluate the probability
of $\vec w$ given  $\{x,n\}$, Alice executes the same transformation
on $\vec w$ evaluating the probability of $\hat O_k \vec w$
by means of Eq.~(\ref{condprob}), that is, the conditional
probability $P(\vec w|x,n,k)$ of $\vec w$ in the extended model
is equal to $P(\hat O_k\vec w|x,n)$, where $P(\cdot|x,n)$
is given by Eq.~(\ref{condprob}).
The unitary transformation performed by Bob guarantees 
that $0\le P(\vec w|x,n,k)\le1$.
Since $\hat O_k\vec v\cdot\hat O_k\vec w=\vec v\cdot\vec w$,
The overall protocol generates the event $\vec w$, given
$\vec v$, with the correct quantum probability 
$(1+\vec v\cdot\vec w)/2$. 

It is worth to note that, before
playing the game, Bob and Alice have to agree about a common
reference frame and the way the Bloch sphere is partitioned. 
This agreement has to be established only at the beginning
and the shared information can be used for any state
and measurement that Bob and Alice wish to test. 
It is also 
interesting to observe that applying this model to the concrete 
example of a $1/2$ spin requires that we enrich the space with a 
preset structure that breaks the rotational symmetry.
This is not surprising, since no (linear or nonlinear) representation 
of three-dimensional rotations exist on a one-dimensional
manifold, that is, it is impossible to represent three-dimensional
rotations by means of differentiable endomorphisms of a one-dimensional
manifold.
The symmetry breaking would occur at the ontological level in
the description of the ontic state, but it would be concealed
at the phenomenological level. However, the ontological shrinking
program does not necessarily imply the spatial symmetry breaking, 
since the dimensional reduction could involve entanglement and not 
single particles. For example, it would be possible in principle to have 
an ontological theory that describes $n$ spins by means of $n$ vectors. 
The ontological space would have a reduced dimensionality, but the 
theory would not break the spatial symmetry. Anyway the ontological
shrinking makes the representation of $SU(N)$ on the ontological space
impossible. 

We have presented the first example of hidden variable model whose 
sampling space dimension is smaller than the quantum state manifold 
one. For example, as previously noted, the Kochen-Specker 
model~\cite{kochen} for a two-state system uses a two-dimensional 
space.

\section{Ontological shrinking for higher dimensional Hilbert spaces}

Our model raises the question whether the 
ontological shrinking is possible also for higher dimensions 
of the Hilbert space. Since in the two-dimensional model
the probability distribution is the mixture of two delta
distributions, a natural generalization of this distribution
would be
\be\label{prob_distr}
\rho(\vec x,n|\psi)=R(n|\psi)\delta[\vec x-\vec f_n(\psi)],
\ee
where $\vec x$ is a $D$-tuple of real variables and $n$
a discrete index that goes from $1$ to $M$. $D$ is the dimension 
of the ontological space. $\vec f_n$ is a generic vectorial function 
and $R(n|\psi)$ is the probability of $n$ given $|\psi\rangle$.
The conditional probability $P(\phi|\vec x,n)$ of an event 
$\phi$ given $\vec x$ and $n$ is such that Eq.~(\ref{born})
is satisfied.

The Beltrametti-Bugajski theory has for example the above
structure with $D$ equal to twice the Hilbert space
dimension and $M=1$~\cite{beltrametti}. In this case there 
is no shrinking since the dimension $D$ of the ontological space
is not smaller than the quantum state manifold dimension. The shrinking
occurs for $D<2 N-2$, $N$ being the Hilbert space dimension.

A simple model with a compressed space and working for a large class 
of events and states has $M=N^2$ and $D=2$. The probability 
distribution $\rho$ and the conditional probabilities for the 
event $\phi$ are
\bey\label{eq1mod}
\rho(X,n,m|\psi)=R(n,m) \delta(X-\psi_n^*\psi_m), \\
\label{eq2mod}
P(\phi|X,n,m)=1-\frac{1}{2R(n,m)}\left|\phi_n^*\phi_m-X\right|^2,
\eey
where $n,m=1,...,N$, $X$ is a complex number and 
$\psi_n\equiv \langle n|\psi\rangle$, $\{|n\rangle\}$ being
a complete set of orthonormal vectors. The distribution
$R(n,m)$ is non-negative and normalized to $1$.
It is easy to check that the functions fulfil the condition
(\ref{born}). However, the conditional probabilities
are positive only if
\be\label{constr}
|\psi_n^*\psi_m-\phi_n^*\phi_m|^2<2 R(n,m).
\ee
It is interesting to note that the selected manifold 
of states $|\psi\rangle$ and events $|\phi\rangle$ have 
the same dimension as the overall quantum state 
manifold. This is a very important property, since
economical ontological models working in a zero measure
region of events or states are quite trivial.

If $R(n,m)$ is constant, it is easy to prove that
the condition 
\be\label{constr2}
|\psi_n-\phi_n|^2<\frac{R}{2}, \text{   for all }n
\ee
is sufficient for positivity. Indeed, from it we have:
\be\begin{array}{l}
|\psi_n^*\psi_m-\phi_n^*\phi_m|^2=    \\
\frac{1}{4}|(\psi_n^*-\phi_n^*)(\psi_m+\phi_m)
+(\psi_n^*+\phi_n^*)(\psi_m-\phi_m)|^2\le  \\
\frac{1}{4}\left(|\psi_n-\phi_n||\psi_m+\phi_m|+|\psi_n+\phi_n||\psi_m-\phi_m|\right)^2\le \\
\left(|\psi_n-\phi_n|+|\psi_m-\phi_m|\right)^2\le 2 R.
\end{array}
\ee

At variance with the previous model for a qubit,
the constraint is not
only on the quantum states, but involves also
the events. Thus, the patching method previously used 
is able to extend the validity of the model
to the whole quantum state space, but not
to the whole set of trace-one projective
measurements. One could try to find the
functions $R(n,m)$ that maximize the
volume of the positivity region, however it is
impossible to cover the whole space of
events. With the choice $R(n,m)=1/N^2$, from
Eq.~(\ref{constr2}) we 
have that a sufficient condition for positivity is
$|\psi_n-\phi_n|\le \frac{1}{\sqrt2 N}$.
This inequality implies that the volume
of a patch region scales at least 
as $(2\pi/N^2)^{N-1}$. Since the volume of the 
whole quantum state manifold scales as 
$\frac{\pi^{N-1}}{(N-2)!}$, the 
additional information required for the patching 
grows as $N \log N$, that is, almost exponentially
with respect the size of the system. 
This choice of $R(n,m)$ is not optimal.
For example, one could give a larger statistical
weight to the events with $n$ or $m$ equal to zero.
If $R(0,n)$ and $R(n,0)$ scale as $1/N$, 
it is easy to show by condition~(\ref{constr}) that both
$\psi_0$ and $\phi_0$ go to $1$ for large $N$.
Using this property, one finds that the constraint~(\ref{constr})
is satisfied if $|\phi_{n\ne0}|\lesssim 1/\sqrt{N}$
and $|\psi_{n\ne0}|\lesssim 1/\sqrt{N}$.
This scaling of the positivity region is a necessary 
condition to have a non-exponential growth of the 
additional information required for the patching.
We will not go into further details on the 
optimization of this model.

There is a simple argument to show that any 
hidden variable theory of the form in 
Eq.~(\ref{prob_distr}) with $M$ finite and working 
for any measurements cannot have an ontological space 
whose dimension is smaller than $2N-3$. Thus, if there 
exists an ontological theory for any measurements that
does not require an exponentially growing
number of resources to describe a single
system, then in this theory $M$
is infinite. The proof is the following.
Using Eq.~(\ref{prob_distr}), Eq.(\ref{born})
becomes
\be\label{PR}
\sum_n P(\phi|\vec x_n,n)R(n|\psi)=
|\langle\phi|\psi\rangle|^2,
\ee
where 
\be\label{eq_xn}
\vec x_n=\vec f_n(\psi). 
\ee
We can assume that there exists a non-zero measure subset
of the quantum state manifold where $R(n|\psi)$ is
different from zero or identically equal to zero
for each $n$. If $R(n|\psi)$ is differentiable almost
everywhere and $M$ finite, it is always possible to find such 
a region. Let us consider in the following only the quantum 
states $|\psi\rangle$ living in this subset. Since the terms in 
Eq.~(\ref{PR}) with $R(n|\psi)\equiv0$ do not contribute, 
we can assume $R(n|\psi)\ne0$ without loss of generality.
The conditional
probability $P(\phi|\vec x_n,n)$ is zero if there exists a 
$|\psi\rangle$ orthogonal to $|\phi\rangle$ such that
$\vec x_n=\vec f_n(\psi)$. If the dimension $D$ of the
ontological space is equal to the quantum state manifold
dimension $2N-2$ and $\vec x_n$ completely identifies
the quantum state $|\psi\rangle$, then the manifold of 
$|\phi\rangle$ where $P(\phi|\vec x_n,n)=0$ is 
$(2N-4)$-dimensional and contains the vectors orthogonal to the one 
vector $|\psi\rangle$ satisfying Eq.~(\ref{eq_xn}).
In the case that one ontological dimension is missed 
(ontological space with $2N-3$ dimensions), one can realize
that the manifold where $P(\phi|\vec x_n,n)$ is
equal to zero has $2N-3$ dimensions. Indeed, the
ontic vector $\vec x_n$ identifies a one-dimensional
manifold of quantum states and the
conditional probability is zero if $|\phi\rangle$ is 
orthogonal to one of these states. For a larger number 
of missed dimensions the manifold with zero probability 
has the same dimension of the overall manifold of $|\phi\rangle$
vectors. 
This means that for an ontological space dimension smaller than
$2N-3$ the overall probability of obtaining $|\phi\rangle$
given $|\psi\rangle$ is zero in a large region
of the events and this region has non-zero measure
if $M$ is finite. But this is impossible since
the probability of $|\phi\rangle$ is 
$|\langle\phi|\psi\rangle|^2$
and is zero only in a zero measure region of
the events. Thus, $M$ cannot be finite.
It is interesting to note that this reasoning does
not forbid the shrinking from $2N-2$ to $2N-3$
for $M$ finite (the model for a qubit gives a practical example),
but it forbids any shrinking in the case of an $N$-dimensional 
Hilbert space with real field.
Indeed in this case the dimension of the manifold of $|\phi\rangle$
orthogonal to $|\psi\rangle$ is $N-2$, where $N$ is the dimension 
of the Hilbert space. Thus, it is sufficient an $(N-2)$-dimensional space,
with only one missed dimension, in order to have $P(\phi|\vec x_n,n)=0$
in a region of events with non-zero measure.
From the above argument it is evident that
the model given by Eqs. (\ref{eq1mod},\ref{eq2mod})
does not work for any measurement, since the probability
distribution contains a finite number of delta 
functions. 

It is important to note that the reported two-dimensional
model is not in contrast with the theorem proved in 
Ref.~\cite{monti2}. That theorem states that the dynamics 
in a theory with dimensional reduction cannot be Markovian.
Indeed, for our model there does not exist a positive 
conditional probability $P_{\hat U}(x,n|\bar x,\bar n)$ 
for every unitary evolution $\hat U$ such that
\be
\rho(x,n|\hat U\psi)=\sum_{\bar n}\int d\bar x
P_{\hat U}(x,n|\bar x,\bar n)\rho(\bar x,\bar n|\psi).
\ee
Indeed, it is possible to prove that the dynamics of the 
probability distribution is not described by a linear
equation, as required for Markov processes.
For this purpose it is sufficient to assume that the Bloch 
vector $\vec v$, defining the quantum state, is in the patch 
$\Omega_1$. This allows us to neglect the additional
index labeling the $M_b$ patches. Thus, the probability
distribution is given by Eq.~(\ref{distr_qubit}) and
lives on the space spanned by $x$ and the binary
discrete variable $n$. Let us consider the Bloch vector 
rotation around the $y$-axis
\be\label{evol_v}
\begin{array}{l}
\frac{\partial v_x}{\partial t}=v_z,  \\
\frac{\partial v_y}{\partial t}=0,   \\
\frac{\partial v_z}{\partial t}=-v_x,
\end{array}
\ee
which correspond in spherical coordinates to
\be\begin{array}{l}
\label{der_sc}
\frac{\partial\varphi}{\partial t}=-\cot\theta\sin\varphi,  \\
\frac{\partial\theta}{\partial t}=\cos\varphi.
\end{array}
\ee

If the dynamics was Markovian, then the time evolution of the
probability distribution $\rho(x,n,t)$ would be described by
the differential equation
\be
\frac{\partial\rho(x,n,t)}{\partial t}=
\sum_{\bar n}\int d\bar x K(x,n|\bar x,\bar n)\rho(\bar x,\bar n,t),
\ee
$K$ being a suitable kernel.

By means of Eqs.~(\ref{distr_qubit}), this becomes
\be\begin{array}{c}
\sin\theta\delta_{n,0}\frac{\partial\delta(x-\varphi)}{\partial\varphi}
\frac{\partial\varphi}{\partial t}+  \\
\left\{\frac{\partial}{\partial\theta}\left[(1-\sin\theta)\delta_{n,1}\delta(x-\theta)
\right]+ 
\cos\theta\delta_{n,0}\delta(x-\varphi) \right\} \frac{\partial\theta}{\partial t}=  \\
K(x,n|\varphi,0)\sin\theta+K(x,n|\theta,1)(1-\sin\theta).
\end{array}
\ee
In particular, for $n=0$, using Eq.~(\ref{der_sc}), we have that
\be\begin{array}{l}
-\cos\theta\frac{\partial\delta(x-\varphi)}{\partial\varphi}
\sin\varphi+\cos\theta\delta(x-\varphi)
\cos\varphi= \\
K(x,0|\varphi,0)\sin\theta+K(x,0|\theta,1)(1-\sin\theta).
\end{array}
\ee
Dividing both sides by $\sin\theta$ and differentiating
with respect to $\theta$, we obtain that
\be\begin{array}{c}
\frac{\partial}{\partial\theta}\left[
-\cot\theta\frac{\partial\delta(x-\varphi)}{\partial\varphi}
\sin\varphi+\cot\theta\delta(x-\varphi)
\cos\varphi\right]=  \\
\frac{\partial}{\partial\theta}
\left[K(x,0|\theta,1)\frac{1-\sin\theta}{\sin\theta}\right].
\end{array}
\ee
There is no $K$ satisfying this equation,
since the left-hand side is a function of both
$\theta$ and $\varphi$, whereas the right-hand side
depends only on $\theta$.
Thus, the dynamical equation of the probability
distribution~(\ref{distr_qubit}) is non-Markovian.

If we allow the dynamics to be non-causal then we should 
consider the possibility that in the preparation-measurement
processes some information could be transferred
back from the measurement apparatus. In other words,
in a possible
extension of the presented models, the probability
distribution could contain a small quantity of information
about $|\phi\rangle$. Thus, a more general question is: what is
the minimal number of continuous variables that the preparation 
and measurement apparatuses have to exchange in order to 
reproduce the quantum probabilities? 

\section{Conclusion}

In conclusion, we have reported a hidden variable
model of measurements where a pure quantum state is 
represented as a statistical mixture of ontic states 
living in a one-dimensional real space. This
model is the first example of ontological shrinking
that works for any state preparation and measurement.
The Wigner~\cite{wigner} and Husimi~\cite{husimi} 
distributions are known examples of statistical representations 
of quantum states on a smaller space. However,
none of them is everywhere non-negative or has
non-negative conditional probabilities associated
with measurements. 
It is interesting to note that, unlike the Wigner and 
Husimi functions, the distribution in our model is
not quadratic in the quantum state. Indeed, any
ontological model of measurement enjoys this
property, as proved in Ref.~\cite{montina}.
Our practical example of shrinking may seem artificial, 
but is important since it raises the interesting question 
of whether a hidden variable description of a quantum
system needs an exponentially growing number of
resources. Furthermore, our model shows that the 
Markov hypothesis is necessary
for the proof of the theorem in Ref.~\cite{monti2}, where we
stated that the ontological space dimension cannot be smaller
than the Hilbert space manifold dimension. In a reversed form,
this theorem states that any hidden variable theory
with an ontological shrinking cannot have a short 
memory dynamics or, in particular, be causal. 
Finally, we have discussed
possible extensions in $N$ dimensions and found
a general property for these models. The possibility
of shrinking considerably the ontological space and the 
introduction of the dynamics in these models are open
questions, whose answer could provide a deeper
explanation of the exponential complexity of quantum 
mechanics. Indeed, the knowledge of the resources
required for a classical simulation of quantum systems
is a very important step for understanding the 
actual computational speed-up of quantum algorithms.

\begin{center}
{\bf Acknowledgments}
\end{center}
\nonumber

I wish to thank K. Wharton,
F. T. Arecchi and L. Hardy for the careful reading of 
the manuscript and valuable comments.
Research at Perimeter Institute for Theoretical Physics is 
supported in part by the Government of Canada through NSERC 
and by the Province of Ontario through MRI.

\end{document}